\documentclass[twocolumn, showpacs, preprintnumbers, amsmath, amssymb, superscriptaddress, pra]{revtex4}
\usepackage{graphicx,amssymb,bm,natbib,amsfonts,amsmath,graphics,color}
\usepackage{bm}
\usepackage{hyperref}

\begin{document}

\title{Far-field emission profiles from L3 photonic crystal cavity modes}

\author{Cristian Bonato}
\affiliation{Huygens Laboratory, Leiden University, P.O. Box 9504, 2300 RA Leiden, the Netherlands}

\author{Jenna Hagemeier}
\affiliation{University of California Santa Barbara, Santa Barbara, California 93106, USA}

\author{Dario Gerace}
\affiliation{Dipartimento di Fisica "Alessandro Volta", Universit\'a di Pavia, via Bassi 6, 27100 Pavia, Italy}

\author{Susanna M. Thon}
\altaffiliation{Current Address: Department of Electrical and Computer Engineering, University of Toronto, 10 King's College Road, Toronto, Ontario M5S 3G4, Canada}
\affiliation{University of California Santa Barbara, Santa Barbara, California 93106, USA}

\author{Hyochul Kim}
\altaffiliation{Current Address: Department of ECE, IREAP, University of Maryland, College Park, Maryland 20742, USA}
\affiliation{University of California Santa Barbara, Santa Barbara, California 93106, USA}

\author{Lucio C. Andreani}
\affiliation{Dipartimento di Fisica "Alessandro Volta", Universit\'a di Pavia, via Bassi 6, 27100 Pavia, Italy}

\author{Pierre Petroff}
\affiliation{University of California Santa Barbara, Santa Barbara, California 93106, USA}

\author{Martin P. van Exter}
\affiliation{Huygens Laboratory, Leiden University, P.O. Box 9504, 2300 RA Leiden, the Netherlands}

\author{Dirk Bouwmeester}
\affiliation{Huygens Laboratory, Leiden University, P.O. Box 9504, 2300 RA Leiden, the Netherlands}
\affiliation{University of California Santa Barbara, Santa Barbara, California 93106, USA}

\begin{abstract}
We experimentally characterize the spatial far-field emission profiles for the two lowest confined modes of a photonic crystal cavity of the L3 type, finding a good agreement with FDTD simulations. We then link the far-field profiles to relevant features of the cavity mode near-fields, using a simple Fabry-Perot resonator model. The effect of disorder on far-field cavity profiles is clarified through comparison between experiments and simulations. These results can be useful for emission engineering from active centers embedded in the cavity.
\end{abstract}

\maketitle

\section{Introduction}

Photonic crystals (PhC) offer unprecedented control over electromagnetic field confinement in all three spatial directions \cite{Joann_book}.
In particular, two-dimensional PhC nanocavities in a planar waveguide have already found applications in different fields such as nanolasers, nonlinear
optics and quantum information processing \cite{yaoLPR10, obrienNP09}. Similar to any electromagnetic resonator,
PhC nanocavity modes are essentially characterized by two figures of merit: the cavity quality factor, $Q$, and the effective confinement
volume of each mode, $V_{\mathrm{mode}}$ \cite{notomiRPP10}. The quality factor is proportional to the photon lifetime in the cavity which depends
on the cavity losses to the external world. The mode volume is a quantitative measure of the spatial confinement of the electromagnetic mode.
In most applications, it is crucial to maximize the $Q/V_{\mathrm{mode}}$ ratio. For example, the Purcell factor, which measures the enhancement
of the spontaneous emission rates for atoms resonant with a cavity is directly proportional to this figure of merit.
In PhC nanocavities the mode is strongly confined to a very small volume, on the order of $(\lambda/n)^3$, where $\lambda$ is the mode
wavelength.
In a planar membrane nanocavity, in-plane confinement is provided by spatial localization of a structural defect in a perfectly periodic PhC with a photonic band-gap, while out-of-plane confinement is given by total internal reflection between the slab and the air cladding
(assuming a suspended membrane as a planar waveguide). Very high quality factors, in the range $10^4$-$10^6$
\cite{akahaneNat03, songNMat05, kuramochiAPL06} have been demonstrated in the literature.
In particular, the L3-type cavity, consisting of three missing holes in a triangular lattice, was the first PhC cavity to show quality
factors larger than $10^4$ \cite{akahaneNat03}.

The spectral mode structure for L3 cavities has been thoroughly investigated by Chalcraft and coworkers \cite{chalcraftAPL07} who compared
the calculated resonant energies, quality factors and emission polarizations for the lowest-order modes with experimental data.
Most experiments coupling single quantum dot emitters to a nanocavity exploit the fundamental (i.e. lowest-energy) cavity mode
\cite{StraufPRL06,HennessyNAT07}.
However, higher order modes can still be important for, e.g., efficient pumping in nanocavity lasers \cite{nomuraAPL06_laser}, selective excitation of
quantum dots embedded within the cavity \cite{nomuraAPL06_QD, oultonOE07}, or mutually coupling quantum dots in different spatial positions \cite{ImamogluJAP07}.  Several groups have studied the near-field emission profiles of photonic crystal nanocavities \cite{mujumdarOE07, intontiPRB08}, even with polarization-resolving imaging \cite{vignoliniAPL09}.

In this paper we report an experimental and theoretical investigation of the spatial far-field profile of the out-of-plane emission for the two lowest-order modes of L3-type PhC nanocavities. We believe the characterization of the out-of-plane far-field emission for PhCs is important for two main reasons.
First, for single-photon source applications the emitted radiation needs to be efficiently collected into a fiber, and simultaneous optimization of far-field
emission for multiple nanocavity resonances could be useful. In addition, in the case of cavity-QED experiments in the ``one-dimensional atom''
approximation, a perfect mode-matching is needed to get a large enough interference between the input light field and the field radiated by the atom
\cite{waksPRL06, poizatPRA07, bonatoPRL10}.
Recently, quite some work has been done to get a beam-like vertical emission from PhC nanocavities \cite{kimPRB06, tranPRB09, toishiOE09, portalupiOE10}
for the fundamental mode. Here we extend previous work by experimentally analyzing the far-field emission properties of both the fundamental and the second-order mode, finding good agreement with numerical simulations. We introduce a simple model, based on a one-dimensional Fabry-Perot resonator, to estimate the essential far-field characteristics of a given near-field mode profile and link them to the relevant device parameters. We believe that such a model can be useful for fast parameter optimization, while full-scale numerical simulations can provide an accurate but time-consuming description of the electromagnetic field in the structure. Finally, we will discuss the effect of fabrication imperfections on the far-field cavity emission profiles. As we will show, measurements of far-field profiles are relatively easy to perform and they can provide insightful information about the parameters and the quality of the cavities under examination.

The paper is organized as follows: in Section II, we present experimental measurements of the far-field profiles and a comparison with theoretical far-fields
extracted from finite-difference-time-domain (FDTD) simulations; in Section III, we introduce a simple model, based on a Fabry-Perot resonator, which is sufficient to give indications of what the actual far-field profile looks like for a given near-field and to link far-field properties to actual device parameters.

\section{Theoretical modeling and experimental data}

\begin{figure}[t]
\centering
\includegraphics[width=7.5 cm] {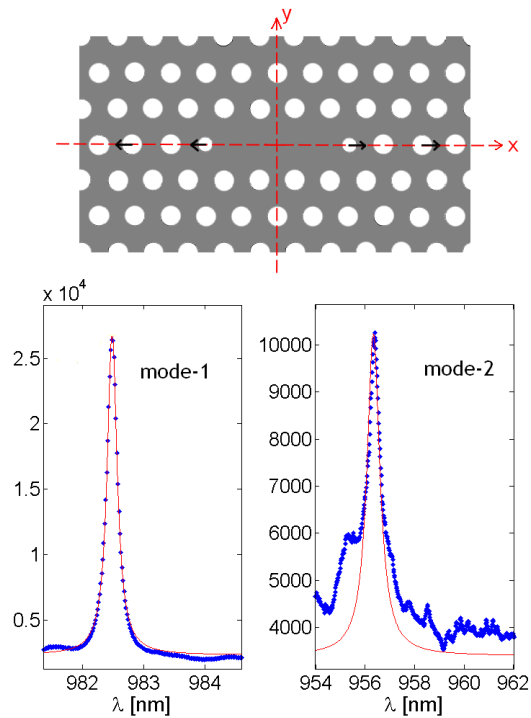}
\caption{Top: sketch of the L3 photonic crystal cavity used in the experiments. Given a design radius $R$ for the holes, the two holes closest to and in-line with the cavity have reduced design radii $R' = 0.75 R$ \cite{andreaniPNN04}. The small holes are shifted from their lattice positions out from the center of the cavity by a fixed quantity
($55$ nm). Finally, the third holes (i.e. two holes away from the small holes) are shifted from their lattice positions out from the center of the cavity by
$45$ nm \cite{akahaneOE05}, to further increase the theoretical $Q$. Bottom: experimentally measured photoluminescence spectra for the two cavity modes.}
\label{fig:spectral}
\end{figure}

Our sample consists of a $180$ nm GaAs membrane grown by molecular beam epitaxy on top of a 0.92 $\mu$m Al$_{0.7}$Ga$_{0.3}$As sacrificial
layer on a GaAs substrate. An In$_{0.4}$Ga$_{0.6}$As quantum dot layer is grown at the center of the GaAs membrane by depositing $10$ periods of
$0.55$ {\AA}-thick InAs and $1.2$ {\AA}-thick In$_{0.13}$Ga$_{0.87}$As. The L3 PhC cavities were fabricated on the sample using standard electron beam
lithography and reactive ion etching techniques \cite{BadolatoScience05, thonAPL09}.
The lattice constant of the triangular hole lattice is $a = 240$ nm. The L3 cavity design was properly modified for $Q$ optimization (see modified holes in Fig. \ref{fig:spectral}) \cite{andreaniPNN04,akahaneOE05}.

The sample was placed in a He-flow cryostat at about $5$ K and illuminated above the GaAs bandgap with a few mW laser beam (wavelength $780$ nm)
on a few $\mu$m$^2$ spot. The photoluminescence from the quantum dot layer embedded in the membrane was collected in the direction normal
to the membrane using a microscope objective with numerical aperture $NA = 0.8$ and spectrally analyzed with a spectrometer (resolution $5.5$ GHz/pixel).
An example of the spectral emission is shown in Fig.~\ref{fig:spectral} for a device with $R = 54$ nm. According to theoretical predictions based on a guided-mode
expansion method~\cite{andreaniPRB06}, the cavity supports two confined modes, with resonances respectively at $\lambda_1^{(th)} = 987$ nm
(theoretical Q-factor $Q_{th} \sim 180000$) and $\lambda_2^{(th)} = 957$ nm ($Q_{th} \sim 15000$). Experimentally, we measured the first-order mode
with a Lorentzian profile centered around $\lambda_1$ = $982.5$ nm with a full-width at half-maximum (FWHM) of $0.195 \pm 0.002$ nm, from which we extract an experimental quality factor of $Q \sim 5000$. On the other hand, the second-order mode has a less perfect Lorentzian lineshape centered around $\lambda_2$ = $956.4$ nm with
FWHM $0.63 \pm 0.03$ nm, and an experimental Q-factor $Q \sim 1500$. Experimental quality factors are lower then the predicted ones due to scattering from fabrication imperfections \cite {portalupiPRB11} and possible absorption from sub-bandgap trap levels and surface states \cite{michaelAPL07}.

\begin{figure}[t]
\centering
\includegraphics[width=7.5 cm] {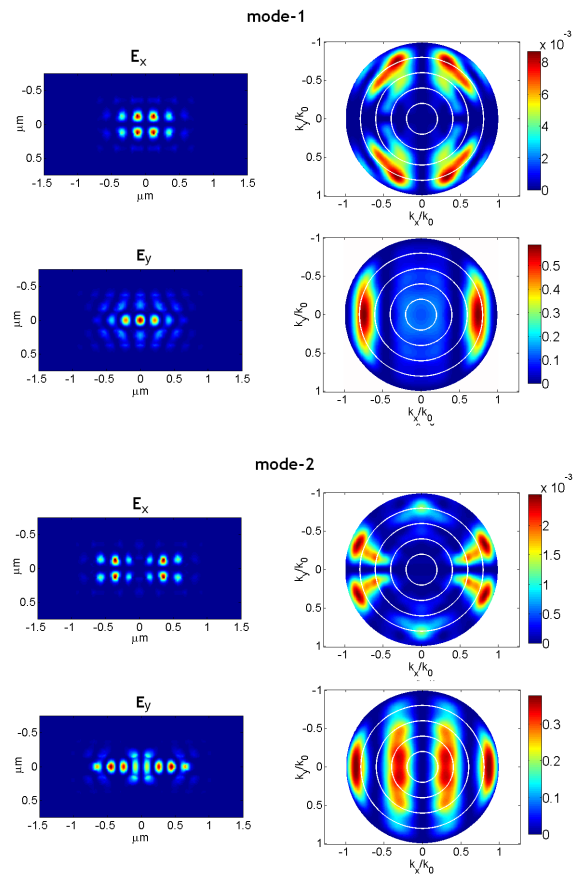}
\caption{FDTD simulations of polarization resolved near-field patterns (left panels) and far-field emission profiles (right panels) for the first and second modes of an L3 cavity. Each mode has two linearly-polarized components. The white concentric rings correspond to a grid with numerical-aperture step of $0.2$. For far-field plots, the color scale is normalized to the totally emitted power in the upper half-space of the PhC cavity (the same normalization factor is used for $E_x$ and $E_y$). For the x-polarization, most of the far-field emission is at large angles, making collection less efficient than for y-polarization and resulting in a strongly polarized photoluminescence signal.}
\label{fig:nearfield}
\end{figure}

Experimentally, the emitted radiation collected from both modes results in a strongly linearly polarized signal.
However, as it is shown in the 3D FDTD simulations of Fig.~\ref{fig:nearfield}, each near-field mode profile has x- and y-components of the electric
field of comparable intensity. The reason for the detection of linearly polarized light can be found by calculating the far-field projections of such
polarization-resolved near-field profiles.\\
The far-field profile can be obtained from the near-field using the procedure introduced by Vuckovic and coworkers \cite{vuckovicIEEE02}.
According to the surface equivalence theorem, all the information about the far-field profile can be obtained from equivalent electric and magnetic
currents, $\mathbf{J}_s = \mathbf{n} \times \mathbf{H}$ and $\mathbf{M}_s = -\mathbf{n} \times \mathbf{E}$,
which depend on the in-plane near field components:
\begin{equation}
\begin{array}{ll}
  N_x = -FT_2 (H_y) & N_y = FT_2 (H_x) \\
  L_x = FT_2 (E_y) & L_y = -FT_2 (E_x)
\end{array}
\end{equation}
where $FT_2$ denoted the two-dimensional Fourier transform.
These equivalent currents are used to calculate the retarded vector potential of the electromagnetic field, which in the far-field can be related
to Fourier transforms of the near-fields. The radiation intensity per unit solid angle can be calculated as:
\begin {equation}
K (\theta, \varphi) = \frac{\eta}{8 \lambda^2} \left( \left| N_{\theta} +\frac{L_{\varphi}}{\eta} \right|^2 +  \left| N_{\varphi} -\frac{L_{\theta}}{\eta} \right|^2 \right)
\label{Eq:FF}
\end{equation}
where $\eta = \sqrt{\mu_0/\varepsilon_0}$ is the impedance of free-space and $\lambda$ is the mode wavelength. The radiation vectors in spherical coordinates can be expressed from their cartesian components as:
\begin  {equation}
\begin{array}{lll}
 N_{\theta} & = & (N_x \cos \varphi + N_y \sin \varphi) \cos \theta \\
 N_{\varphi} & = & -N_x \sin \varphi + N_y \cos \varphi \, \, ,
\end{array}
\end{equation}
and similarly for $\mathbf{L}$.
The far-field profiles calculated from the near-fields in the left panels of Fig.~\ref{fig:nearfield} are shown in the same figure, on the right.
In these plots, the color scale is normalized to the totally emitted power in the upper half-space of the PhC cavity (the same normalization factor is used for $E_x$ and $E_y$).
Most of the emission from the x-polarization of both modes is predicted at very large angles,and therefore is inefficiently collected
by commonly employed microscope objectives. This results in the strong linear polarization observed in the photoluminescence spectra.

To perform a direct measurement of the far-field emitted intensity at each resonant mode frequency, the filtered photoluminescence at the back focal plane of the microscope objective is imaged. Given a characteristic size of the near-field emission, $p \sim $ 500 nm, at a wavelength $\lambda \sim 1$ $\mu$m the Fresnel number is
$F = p^2/(L \lambda) \sim 0.01$, well in the far-field regime ($L \sim 2$ mm).
The far-field was imaged on an intensified CCD camera by a lens with focal length $40$ cm in a $2f-2f$ configuration. To make sure that we
were looking at the microscope objective back focal plane, we adjusted the lens to see the sharp image of the objective edge on the CCD.
This sharp edge was used to calibrate the numerical aperture scale of the far-field images, assuming that the sharp edges correspond to
the $NA = 0.8$ of the objective employed. An interference filter, with a bandwidth $1$ nm, was used to spectrally select the mode of interest. Images were collected after integrating for $30$ s and the background noise was removed by subtracting an image taken with a slightly tilted interference filter.

\begin{figure}[t]
\centering
\includegraphics[width=7.5 cm] {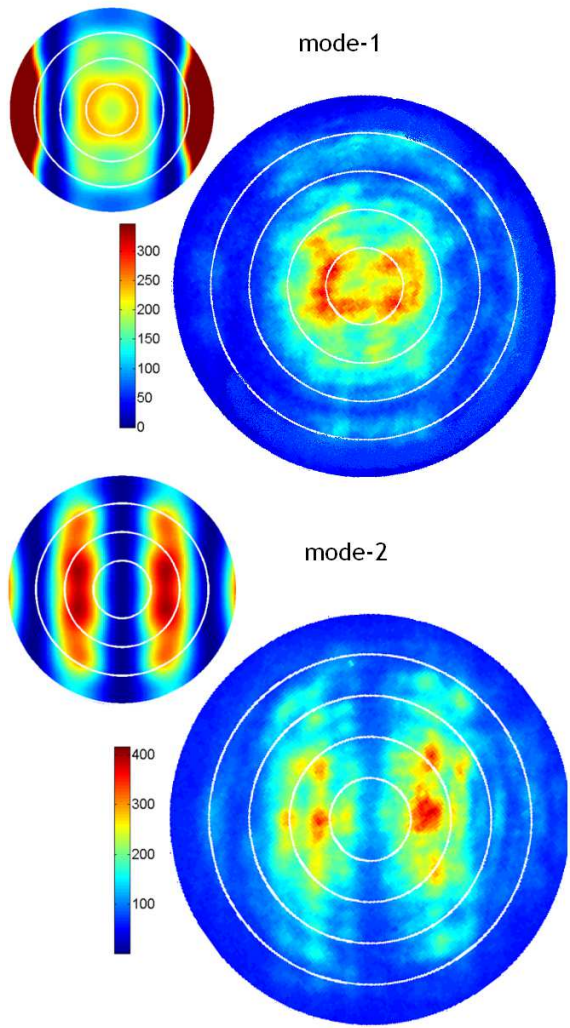}
\caption{Experimental far-field emission profiles for the two L3 cavity modes (larger panels) and the corresponding predictions by FDTD simulations (smaller panels on the top left of the experimental plots), resulting from the sum of x- and y-components in Fig. 2. The white concentric rings correspond to a grid with numerical-aperture step of $0.2$.
The color scale bars show the detected photon counts per second. The counts for the two profiles cannot be compared since they depend not only on the device parameters, but also on the density of quantum dots emitting in the wavelength range of the specific cavity mode.}
\label{fig:farField}
\end{figure}

The experimental far-field spatial emission profiles for the first-order and second-order modes are shown in the two larger plots on the right side of
Fig.~\ref{fig:farField}, together with the far-field projections obtained from FDTD simulations (smaller insets on the left) for a direct comparison.
The first-order mode exhibits a centrally illuminated area extending to about $NA \sim 0.5$, with a ring-like structure inside ($NA \sim 0.2$),
matching the low-NA portion of the simulated far-field. The simulated far-field suggests that most of the light from the first-order mode is emitted in two high-NA lobes, which are not collected at all by our set-up.
The far-field profile for the second-order mode consists of two lobes, whose center is at a minimal $NA \sim 0.3$.\\

\begin{figure}[t]
\centering
\includegraphics[width=7.5 cm] {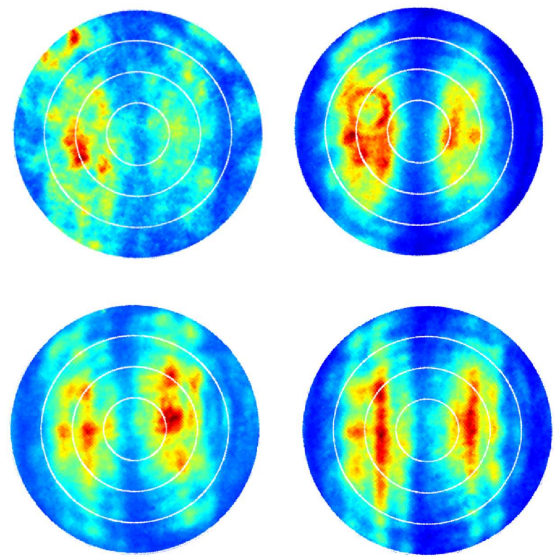}
\caption{Experimental far-field profiles of the second-order mode emission from four different devices on the same wafer. Finer details are reproducible in different measurements performed on the same device, but differ from device to device. The white concentric rings correspond to a grid with numerical-aperture step of $0.2$.}
\label{fig:imperf}
\end{figure}

\begin{figure}[t]
\centering
\includegraphics[width=7.5 cm] {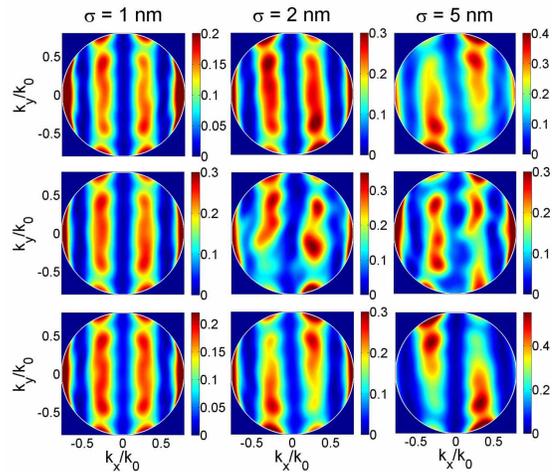}
\caption{Far-field intensity profiles for varying degrees of disorder, normalized by the total power radiated into the upper half space. The images are cut-off at a numerical aperture value of $0.8$, indicated by the white line, to allow for better comparison with experimental data. Data are shown for disorder parameters of $\sigma = 1$ nm (left column), $2$ nm (middle column), and $5$ nm (right column).}
\label{fig:disorder1}
\end{figure}

Finer details also appear inside the two lobes, in the form of spots separated by $\Delta k /k_0 \sim 0.1 - 0.2$.
Such structures correspond, in the near-field, to light sources which are separated from the cavity about $5-10$ times the characteristic size
of the cavity mode. The simulated near-field profiles shown in Fig.~\ref{fig:nearfield} suggest that the optical field decays very fast out of the cavity region, implying that such features might be due to light that escapes from the cavity due to fabrication imperfections \cite{portalupiPRB11}. The fine details are reproducible for different measurements performed on the same device. Finally, we show in Fig. \ref{fig:imperf} the far-field profiles of the second-order modes for different devices on the same wafer.
Each plot shows the characteristic two-lobes profile, as expected for this mode, with reproducible finer details that appear to be device-dependent.

To investigate the cause of fine structure within the two-lobe far-field pattern of the second order mode, FDTD simulations were done in which disorder was introduced. It is assumed that, due to fabrication imperfections, the dominant type of disorder is in the hole radii of the PhC lattice. Therefore, the hole radii $R$ were varied randomly according to the distribution function $P(R)\propto\exp[-(R-\bar{R})^2/2\sigma^2)]$. Some simulation data are given in Figure~\ref{fig:disorder1}, showing disorder introduced to the far-field profiles as a result of increasing the disorder parameter $\sigma$. These results indicate that far-field measurements could be used as an indicator of disorder in the lattice structure.

\section{A simple Fabry-Perot model}

In general, 3D FDTD simulations can provide accurate modeling of near-field and far-field properties of PhC cavity modes. However, they give little physical insight on how the detected features of such far-field profiles can be related to specific device parameters.
In this Section, we will show that the experimental data can be reproduced by a simple model, elaborated from the proposal of Sauvan
and coworkers \cite{sauvanPRB05}.

For a line of $N \geq 3$ missing holes, the PhC nanocavity can be described quite accurately by a Fabry-Perot
resonator, in which the fundamental Bloch mode of a single-line-defect PhC waveguide is trapped between two PhC mirrors of modal
reflectivity $r(\lambda)$. The properties of such a cavity are shown to depend only on three parameters, namely the group index $n_g$ of the
Bloch mode, the reflection coefficient $r(\lambda)$ of the mirrors and the effective cavity length $L$.
The Bloch mode can be calculated as the eigenstate of the PhC waveguide in the Fourier basis, and its modal reflectivity can be obtained
with the method described in Ref.~\cite{silbersteinJOSAA01}. Fabry-Perot models have been shown to be a useful tool to probe cavity resonances and the group index of photonic crystal waveguides \cite{combrieOE06, lalanneLPR08}, and to describe acousto-mechanical cavity tuning effects \cite{fuhrmannNatPhot11}.
We consider the $E_y$ near-field profiles shown in Fig.~\ref{fig:nearfield}. Taking the intensity distribution along the $y = 0$ axis, the modes
show a sinusoidal intensity distribution with nodes and antinodes in the cavity region, decaying exponentially outside the cavity region.
From simulations, the intensity decay length for the first-order mode is $\delta_1 \sim 220 \pm 10$ nm, while it is $\delta_2 \sim 155 \pm 8$ nm
for the second-order mode. This is very similar to the intensity distribution in a cavity between two Distributed Bragg Reflectors (DBR).
For the $E_y$ near-field profiles most of the radiation is emitted along the $y = 0$ axis, with little structure and weaker intensity outside.
Given the relevance of such polarization in determining the far-field emission properties, to a first-order approximation it makes sense
to consider a one-dimensional model, which takes into account only the structure along the $y=0$ axis.
Such a model has the advantage of being extremely simple, although able to give significant hints on the main far-field
profile properties. A two-dimensional model based on an effective index approximation should be used (with no significant difficulties)
to study also the $E_x$ profiles.

\begin{figure}[t]
\centering
\includegraphics[width=7.5 cm] {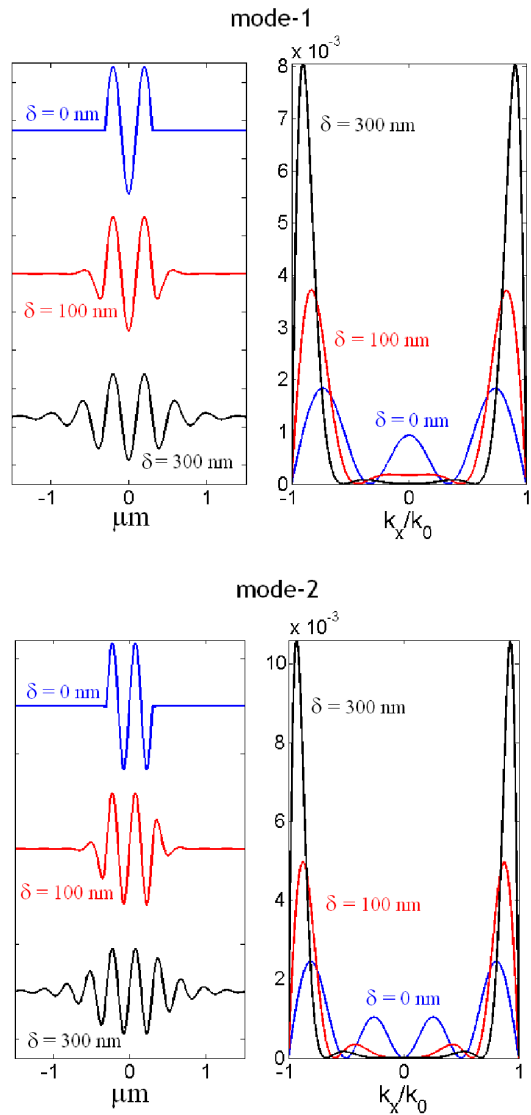}
\caption{Far-field results for the one-dimensional Fabry-Perot model (first and second-order modes). For each mode, on the left side the expected near-field profile
considering a Fabry-Perot cavity with DBR mirrors with varying penetration depth $\delta$.
On the right side, the corresponding far-field profiles calculated from Eq.~\ref{Eq:Kmodel}. For the first-order mode, in the region $|k_x|/k_0 < 0.5$,
there is a single peak centered at $k_x/k_0 = 0$ for small penetration depth. This peak flattens and broadens on increasing penetration depth.
For large enough penetration depth, the central peak splits into two smaller peaks. The second-order mode, in the region $|k_x|/k_0 < 0.5$,
consists of two peaks that split further and further out for increasing penetration depth $\delta$. }
\label{fig:FP_1D}
\end{figure}

Let us consider a one-dimensional Fabry-Perot resonator, with the field intensity profile varying sinusoidally in the cavity region and exponentially
decaying in the mirror regions, with characteristic penetration depth $\delta$.
The resonant frequency $\Omega_m$ can be calculated by imposing the total phase acquired by the Bloch mode traveling back and forth to be a
multiple of $\pi$. The far-field profile can be calculated using the procedure outlined in the previous Section (Eq.~\ref{Eq:FF}).
Since the modes are TE-like modes, only $E_x$, $E_y$ and $H_z$ are non-negligible and only the $L_x$ component is relevant in Eq.~\ref{Eq:FF}.
Let's consider a separable electric field distribution $E_y (x, y) = \alpha (x) \beta(y)$. The component $L_x$ is separable as well:
$L_x = \tilde{\alpha} (k_x) \tilde{\beta}(k_y)$, where $\tilde{\alpha}$ and $\tilde{\beta}$ are the one-dimensional Fourier transforms of $\alpha (x)$
and $\beta(y)$, making:
\begin {equation}
\begin{array}{c}
 K (\theta, \varphi) \propto |\tilde{\alpha} (k_0 \sin\theta \cos\varphi) |^2 |\tilde{\beta} (k_0 \sin\theta \sin\varphi) |^2 \times\\
  \qquad \qquad  \times \left( \cos^2\varphi \cos^2\theta + \sin^2\varphi  \right)
\end{array}
\label{Eq:Kmodel}
\end{equation}
In a simplified one-dimensional model, $\beta (y)$ is narrow in real-space, so its Fourier transform is wide and can be neglected ($|\tilde{\beta}(k_y)|^2 \sim 1$).
If we look at the far-field distribution along the x-axis we select $\varphi = 0$, so that the resulting one-dimensional far-field profile is:
\begin {equation}
K(\theta) \sim  |\tilde{\alpha} (k_0 \sin\theta)|^2 \cos^2\theta
\end{equation}
or, in terms of transverse wavevectors: $K (k_x) \sim |\tilde{\alpha} (k_x)|^2 (k_0^2 - k_x^2)$.

\begin{figure}[t]
\centering
\includegraphics[width=7.5 cm] {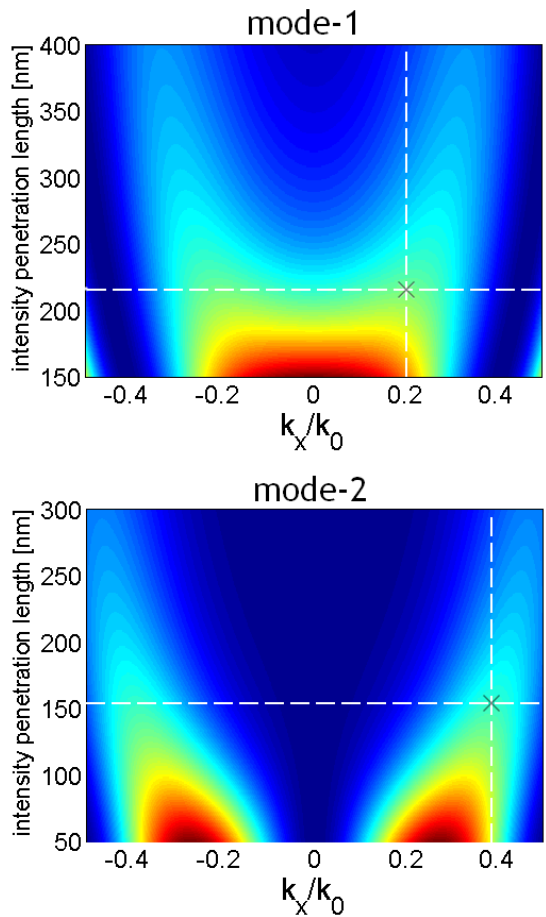}
\caption{One-dimensional far-field profiles as a function of the intensity penetration depths calculated using the simple Fabry-Perot
model for the first (left-side plot) and the second (right-side plot) order modes.
The white dashed lines correspond to the experimental position of the center of the lobes (Fig. \ref{fig:farField}) and the penetration
depth from FDTD simulations.}
\label{fig:penDepth}
\end{figure}

An analytical solution can be found and is reported in Appendix A. However, the resulting formula is too complicated to give intuitive insights; therefore we will just discuss
the numerical results (from Eq. \ref{Eq:theory}) in this context. Fig.~\ref{fig:FP_1D} shows the expected far-field profiles for the first and second-order modes for different values of the penetration depth, $\delta$. The first-order mode exhibits a central peak (centered at $k = 0$) and two outer lobes as predicted by the FDTD simulations.
We see that, for increasing penetration depth, the two main outer lobes become narrower in k-space and more outward, while the central peak broadens and flattens.
A similar behavior can be observed in the second-order mode. Here, the far-field profile is given by two central peaks and two outer lobes, and again, for increasing penetration depth $\delta$ the two outer lobes move outwards and become more localized.

The Fabry-Perot resonator model was shown to give results for the quality factor $Q$ in agreement with more sophisticated FDTD
simulations \cite{sauvanPRB05}:
\begin {equation}
Q = \frac{k_0}{1-R} n_g \left( L + 2 L_p \right)
\end{equation}
where the group delay $\tau = \frac{\partial \phi_r}{\partial \omega} $ experienced by the light upon mirror reflection enters as a
characteristic length $L_p = -\lambda_0^2/(4 \pi n_g) (\partial \phi_r/\partial \lambda)_{\lambda_0}$.
In general, for PhC structures (for example, heterostructure mirrors) it has been shown that $L_p$ can be unrelated to the
characteristic damping length of the energy distribution inside the mirrors $\delta$ \cite{sauvanAPL09}.
However, in the configuration under investigation, the classical relation for a DBR, $L_p \leq \delta$, is valid, with the equality
being strictly fulfilled only in the limiting case of quarter-wave mirrors with low refractive index contrast, $\Delta n/n$
($L_p = \delta = n \Lambda /(2 \Delta n)$).

In Fig. \ref{fig:penDepth}, the emission in the region $|k_x/k_0|<0.5$  is shown as a function of the penetration depth.
In the case of the fundamental mode, there is a single peak centered around $k = 0$ for small penetration depth, with
a quite flat profile and broadening for $|k_x/k_0| < 0.25$.
When the penetration depth increases beyond $\delta \sim 150$ nm, the far-field emission splits into two lobes, which
separate more and more on increasing $\delta$.
At $\delta \sim 250$ nm, the peaks are centered around $k \sim \pm 0.25 k_0$, corresponding to an opening angle of
about $15$ degrees. These findings explain the ring-like structure we observe in the experimental far-field: the vertical
dashed white line in the figure shows the experimental central NA for the peaks (from Fig. \ref{fig:farField}), which corresponds
to a penetration depth of about $220$ nm, fully compatible with the penetration depth from FDTD simulations, $\delta_1$.
For the second-order mode, in the region $|k_x/k_0|<0.5$ there are always two emission lobes, even at small
penetration depth. On increasing $\delta$, the two lobes move further and further apart.
From Fig.~\ref{fig:farField}, the lobes are centered around $NA \sim 0.35$, which in our simulations is consistent with a
penetration depth of about $150$ nm, as predicted by FDTD simulations.
Therefore, far-field profiles can provide useful information about the penetration depth $\delta$ of PhC cavity modes,
which in this case offer a bound on the effective cavity length, $L_p$, and therefore on the Q-factor.
Extensions of the present analysis to treat coupled cavity modes~\cite{intontiPRL, brunsteinAPL11} can also be considered.

\section{Conclusions}

In conclusion, we have presented an extensive characterization of far-field emission profiles from L3-type photonic crystal nanocavities,
introducing a simple imaging technique as an efficient tool to give a two-dimensional mapping of the emitted intensity.
The measurements have been directly compared to theoretically modeled far-field projections from the 3D FDTD near-field
cavity modes profiles, and we believe these results to be useful to PhC cavity designs for specific purposes. The effect of disorder on the far-field profiles was investigated via numerical simulations.\\
Finally,we have introduced a simple Fabry-Perot model that is able to capture the essential features of far-field properties
for suitably designed near-field profiles. As a particular application of this model, we can envision, for example, the simultaneous optimization of  in- and out-coupling  for two different
modes supported by the same PhC cavity, which is still an open problem that might benefit from simplified models like the one presented here.

\section*{Acknowledgments}
The authors acknowledge helpful discussions with Morten Bakker. This work was supported by NSF NIRT Grant No. 0304678, Marie Curie EXT-CT-2006-042580 and FOM$\backslash$NWO grant No. 09PR2721-2.  A portion of this work was done in the UCSB nanofabrication facility, part of the NSF funded NNIN network.

\section*{Appendix A: Analytical calculations}
The resonance frequencies for the modes can be found by setting the condition that the total phase acquired by the Bloch mode traveling back and forth is a multiple of $\pi$ ($2 k_m L = m \pi$), which results in:
\begin {equation}
k_m = \frac{m \pi}{2L}
\end{equation}
Let us start with a perfectly confined mode, with no penetration into the mirrors ($\delta = 0$). In this case the field is given by:
\begin{equation}
\alpha_{\delta=0} (x) = \Pi \left(\frac{x}{L}\right) e^{i(k_m x +\phi_m)} + \mbox{c. c.}
\end{equation}
where $\Pi (x)$ is the rectangular function, $\Pi (x) = 1$ for $|x|<1/2$ and zero elsewhere. The phase $\phi_m$ is set by the boundary conditions: $\phi_m = 0$ for $m$ even (cosine-like solutions) and  $\phi_m = \pi/2$ for $m$ odd (sine-like solutions). From Eq. \ref{Eq:Kmodel}:

\begin{equation}
\begin {array} {ll}
 K_1 (\theta) & \sim \left| e^{i\phi_m}\mbox{Sinc} \left[ \frac{m \pi}{2} \left( \sin\theta-1  \right)\right] + \right. \\
   & \qquad + e^{-i\phi_m} \left. \mbox{Sinc} \left[ \frac{m \pi}{2} \left( \sin\theta+1  \right)  \right]  \right|^2 \cos^2\theta
  \end{array}
\label{Eq:K1_Sinc}
\end{equation}

The two modulo-squared Sinc functions in Eq. \ref{Eq:K1_Sinc} give two main peaks centered at $\theta = \pm \pi/2$, which correspond to the higher-NA peaks in the far-field in Fig. 6. The first relative maximum of $\mbox{Sinc}^2(x)$ is at $x \sim 1.4303\pi$, which for Eq. \ref{Eq:K1_Sinc} corresponds to $\sin\theta \sim  \pm (1-2.861/m)$. The FWHM of such a peak for $\mbox{Sinc}^2(x)$ is $0.522\pi$, which corresponds to $\delta k_x \sim 1.044/m$ ($\delta k_x \sim 0.35$ for the first-order mode, corresponding to $m=3$, and $\delta k_x \sim 0.26$ for the second-order mode, corresponding to $m=4$). Therefore for the first-order mode, the first relative maxima of the \mbox{Sinc} functions superpose, giving just one central peak. For the secod-order mode, on the other hand, the first relative maxima are well separated.

Including the penetration depth $\delta$ into the model, a simple near-field profile can be taken as a superposition of $\Pi \left(\frac{x}{L}\right)$ and two exponentially-decaying wings, as follows:
\begin{equation}
\begin {array} {ll}
\alpha (x) &= \left[ \Pi \left(\frac{x}{L}\right) + H \left( x - \frac{L}{2} \right) e^{-(x-L/2)/\delta} \right. + \\
& \qquad \left. + H \left( -x - \frac{L}{2} \right) e^{-(-x-L/2)/\delta} \right] e^{i(k_m x +\phi_m)}  + \mbox{c. c.}
\end{array}
\end{equation}
where $H(x)$ is the Heaviside function ($H(x) = 1$ for $x>0$, $H(x) = 0$ for $x<0$). The part around $x=0$ ($\Pi \left(\frac{x}{L}\right)$) gives the same Fourier-transform as in Eq. \ref{Eq:K1_Sinc} (which we label $\tilde{\alpha}_1 (\sin\theta)$)), while the left and right-side exponential decay regions give the following:
\begin {equation}
\begin{array}{lll}
\tilde{\alpha}_2 (\sin\theta) &\sim& e^{i\phi_m} \frac{e^{i m (\pi/2) (\sin\theta + 1)}}{\delta/L - i m \pi (\sin\theta + 1)} + e^{-i\phi_m} \frac{e^{i m (\pi/2) (\sin\theta - 1)}}{\delta/L - i m \pi (\sin\theta - 1)}\\
\\
\tilde{\alpha}_3 (\sin\theta) &\sim& e^{i\phi_m} \frac{e^{-i m (\pi/2) (\sin\theta \pm 1)}}{\delta/L + i m \pi (\sin\theta \pm 1)} + e^{-i\phi_m} \frac{e^{-i m (\pi/2) (\sin\theta - 1)}}{\delta/L + i m \pi (\sin\theta - 1)}
\end{array}
\end{equation}
All the quantities depend on the ratio $\delta/L$ between the penetration depth and the cavity length. The far-field profile can be calculated to be:
\begin {equation}
K_1 (\theta) = \left| \sum_{j=1}^3 \tilde{\alpha}_j (\sin\theta)  \right|^2 \cos^2\theta
\label {Eq:theory}
\end{equation}

\bibliographystyle{model1-num-names}
\bibliography{phCrystals}

\end{document}